\documentclass[%
 aip,
 amsmath,amssymb,
 reprint,%
]{revtex4-1}

\usepackage{graphicx}
\usepackage{dcolumn}
\usepackage{bm}

\usepackage[utf8]{inputenc}
\usepackage[T1]{fontenc}
\usepackage{mathptmx}
\usepackage{etoolbox}

\makeatletter
\def\@email#1#2{%
 \endgroup
 \patchcmd{\titleblock@produce}
  {\frontmatter@RRAPformat}
  {\frontmatter@RRAPformat{\produce@RRAP{*#1\href{mailto:#2}{#2}}}\frontmatter@RRAPformat}
  {}{}
}%
\makeatother
\begin{document}

\preprint{AIP/123-QED}

\title{Observation of breakdown wave mechanism in avalanche ionization produced atmospheric plasma generated by a picosecond CO$_2$ laser}
\author{E. Welch}
 \email{ewelch@ucla.edu}
\author{D. Matteo}
\author{S. Tochitsky}
\author{G. Louwrens}
\author{C. Joshi}
\affiliation{Neptune Laboratory, University of California at Los Angeles, Los Angeles, CA 90095 USA}

\date{\today}

\begin{abstract}
Understanding the formation and long-timescale evolution of atmospheric plasmas produced by ultrashort, long wavelength IR (LWIR) pulses is an important but partially understood problem. Of particular interest are plasmas produced in air with a peak laser intensity $\sim$10$^{12}$ W/cm$^2$, the so-called clamping intensity observed in LWIR atmospheric guiding experiments where tunneling and multi-photon ionization operative at near-IR or shorter wavelengths are inoperative. We find that avalanche breakdown on the surface of aerosol (dust) particles can act to seed the breakdown of air observed above the 200 GW/cm$^2$ threshold when a train of 3 ps 10.6 $\mu$m laser pulses separated by 18 ps are used. The breakdown first appears at the best focus but propagates backwards towards the focusing optic as the plasma density approaches critical density and makes forward propagation impossible. The velocity of the backward propagating breakdown can be as high as 10$^9$ cm/s, an order of magnitude greater than measured with ns pulse-produced breakdown and can be explained rather well by the so-called breakdown wave mechanism. Transverse plasma expansion with a similar velocity is assisted by UV photoionization and is observed as a secondary longitudinal breakdown mechanism in roughly 10 percent of the shots. When a cm size, TW power beam is propagated, interception of aerosol particles is guaranteed and several (40 cm$^{-3}$) breakdown sites appear, each initially producing a near critical density plasma. On a 10 ns-1 $\mu$s timescale, shockwaves from each site expand radially and coalesce to produce a large hot gas channel. The radial velocity of the expansion agrees well with the prediction of the blast wave theory developed for ultrafast atmospheric detonations.
\end{abstract}

\maketitle

\section{\label{sec1}Introduction}
The advent of MW-power, Q-switched (laser wavelength, $\lambda\sim$1 $\mu$m) or naturally gain switched ($\lambda\sim$10 $\mu$m), nanosecond pulsed lasers in the early 1960s resulted in the observation of laser-induced breakdown of air. Laser spark or gas ionization was initiated near the focus of the laser beam. The measured breakdown threshold for nanosecond pulses was $\sim$10$^9$ and $\sim$10$^{11}$ W/cm$^2$ for 10.6 $\mu$m and 1.06 $\mu$m \cite{Smi70}, respectively, and decreased inversely with the focal spot size diameter \cite{Smi71}.  The recorded wavelength differences in the breakdown threshold were associated with much more efficient avalanche ionization at long wavelengths \cite{Yab74a}. The avalanche ionization process relies on collisional cascading of the seed electrons which in turn are produced on aerosol (or dust) particles at laser intensities much lower than the breakdown threshold of pure gases \cite{Yab74a,Len73,Smi75}. Once the plasma electron density increases sufficiently so that plasma begins to emit free-free (bremsstrahlung) and free-bound (recombination) radiation, a spark can be readily observed usually at the focal point of the laser. At this stage of optical breakdown, the gas is locally often fully ionized and the laser pulse is strongly attenuated and typically cannot propagate downstream of the breakdown. However, the plasma appears to travel at several kilometers per second longitudinally towards the focusing optic \cite{Ram64,Alc68a,Shi19} and expand transversely in the unionized gas. Various mechanisms of energy transfer from this optical discharge plasma to adjacent cold gas have been considered for explaining the velocity of backward (towards the laser) plasma propagation in air. In increasing velocity order, these mechanisms are as follows: thermal conduction (optical combustion), shockwave propagation (optical detonation), ionization by UV radiation from the plasma (fast ionization wave), and the breakdown wave mechanism. They are all described in detail by Y.P. Raizer in Ref.~\onlinecite{Rai77}.\par
Recent demonstrations of laser guiding in air without significant expansion over tens of vacuum Rayleigh lengths---often termed air filamentation---when femtosecond and picosecond long laser pulses have power exceeding the critical power for Kerr self-focusing has renewed interest in understanding the role of laser-induced collisional ionization in optical breakdown phenomena. This is because for such long distances self-guiding due to Kerr self-focusing is thought to be countered by laser produced plasma defocusing. So it is imperative to understand the mechanism(s) of ionization and the rate at which the plasma density increases during the passage of the laser pulse across a given position in space. For ultrashort (less than $\sim$100 fs) optical pulses, the avalanche process does not have sufficient time to fully ionize atmospheric air and therefore any visible-to-eye plasma is not produced by avalanche ionization. In filaments generated by such ultrashort near IR and visible laser pulses, the electron density of the order of 10$^{16}$ cm$^{-3}$ is thought to be generated by optical field ionization via a combination of tunnel and multi-photon ionization (MPI) processes \cite{Fon99,Cou07}. This field ionization depends on the Keldysh parameter $\lambda_k=\sqrt{I_p/2U_p}$, where $I_p$ is the ionization potential and $U_p$ is the laser ponderomotive energy \cite{Kel65}. The familiar limiting cases are MPI for $\lambda_k$>>1 and tunnel ionization for $\lambda_k$<<1. The situation changes when self-guiding occurs with picosecond 10.6 $\mu$m pulses for which the MPI process becomes very inefficient in ionizing air (e.g. O ionization potential is 12.08 eV for the 2p$^4$ electron which would necessitate the simultaneous absorption of $\sim$100 photons to ionize this outermost electron by MPI).\par
In the 1970s, CO$_2$ lasers were first mode-locked to generate nanosecond (ns) 10 $\mu$m pulses \cite{Woo70} and subsequently the free-induction decay technique \cite{Yab74b} was used to generate sub-ns pulses. It was not until 1990s that picosecond CO$_2$ laser pulses \cite{Cor85,Toc99} were routinely produced and amplified in high pressure amplifiers \cite{Hab10} to give TW peak power laser pulses. Such high-power CO$_2$ pulses made it possible to study guiding of long-wave infrared (LWIR) pulses in air. Recent simulations have suggested that aerosols may assist in the self-guiding of LWIR pulses through the air \cite{Woo20}. Also, for such a CO$_2$ laser filament in the atmosphere, the measured average clamped intensity was $\sim$1 TW/cm$^2$, more than an order of magnitude smaller than the so called barrier suppression threshold of O$_2$/N$_2$ \cite{Toc19,Wal94} where the tunnel ionization probability becomes unity. Note that by definition the plasma-density in such air filaments must be far below the critical density for the laser pulse that is being guided. The question is what is the maximum free electron density that exists in the filament and how is it produced? Indeed, in the 10 $\mu$m, picosecond pulse filamentation experiment reported in Ref.~\onlinecite{Toc19}, no full ionization of air related to the optical breakdown was claimed-as was reported in earlier $\sim$100 ns CO$_2$ laser atmospheric breakdown experiments \cite{Gre78,Shi93,Aut81}. To explain picosecond 10 $\mu$m filamentation, it is therefore imperative to understand the role of avalanche or collisional ionization as a process that can generate a sufficient number of free carriers in air required for plasma defocusing to balance the Kerr self-focusing and create conditions for guided beam propagation. The above discussion highlights a need to understand how rapidly plasma formation occurs and the role avalanche ionization plays in the plasma expansion in air. To carry out such a study, one needs a picosecond high power long wavelength laser. In a broader context of physics of optical breakdown of gases, it is not clear to what extent the mechanisms of air-plasma expansion described in the literature \cite{Rai77} are applicable for short pulse high-power lasers with intensities several orders of magnitude higher than that typically realized for nanosecond pulses. However, the avalanche assisted air breakdown threshold (\textit{laser intensity required to generate a plasma spark that is observable with the naked eye}) and plasma dynamics at intensities $\sim$10$^{12}$ W/cm$^2$ where optical guiding occurs are not understood for a picosecond 10 $\mu$m laser pulse.\par
In this paper we study optical breakdown by using GW and TW power, nominally 3 ps (FWHM) CO$_2$ laser pulses focused in laboratory air to $\sim$10$^{12}$ W/cm$^2$ intensities. In order to measure the onset of plasma formation and its subsequent expansion we apply two different interferometry techniques using a visible probe beam. To determine the contribution of aerosol particles within the ionized gas volume we probe two focusing geometries producing a small ($\sim$100 $\mu$m diameter) focal volume where the probability of encountering an aerosol particle can be ignored, and a large $\sim$1 cm diameter spot that has a long Rayleigh range where the laser pulse routinely encounters several particles that ionize readily to seed the plasma formation at a similar level of laser intensity. In either case the peak intensity of the laser pulse is kept at 10$^{12}$ W/cm$^2$.\par
The paper is organized as follows. In Sec.~\ref{sec3} we review several different physical mechanisms that have been invoked to explain the observed characteristics of plasmas produced during breakdown of gases using laser pulses in the past. Most of the experimental work was carried out using ns class laser pulses. Since we are interested in extending these studies to ps, long-wavelength regime we have developed a ps MOPA CO$_2$ laser system that can generate TW peak power pulses. For completeness, in Sec.~\ref{subsec4.1} we briefly describe this high-power CO$_2$ laser system at the UCLA Neptune Laboratory and its parameters. In Sec.~\ref{subsec4.2} we discuss results of space and time resolved measurements in air-plasma generated by a tightly focused CO$_2$ laser beam of a GW power. Here, better than 10 ps time resolution provided by a streak camera diagnostic allows the recording of fast plasma dynamics in a time interval from 0.01 to 2 ns. We also analyzed 2D distribution of plasma density on longer time scales using Michelson interferometry. In Sec.~\ref{subsec4.3} we present the results of interferometry in plasmas generated by a large, cm-diameter weakly-focused TW CO$_2$ laser beam. By using a 10 ns green probe pulse, we measure 2D dynamics of breakdown plasma at a longer timescale up to $\sim$1 $\mu$s, close to the near complete recombination of the plasma. In Sec.~\ref{sec5} we summarize our findings.
\section{\label{sec3}Laser-induced avalanche breakdown in air and its upstream propagation towards the laser---Theory}
In 2017, self-guiding of 2 ps CO$_2$ laser pulse was observed when a multi-Joule laser pulse was focused in air \cite{Toc19}. The laser eventually came to an equilibrium spot size of 1 cm that contained about half the initial laser energy (intensity of $\sim$10$^{12}$ W/cm$^2$) and propagated for more than 30 meters without significant spreading. Although no plasma was visible in this long channel, simulations indicated that defocusing provided by a low density ($\sim$10$^{14}$ cm$^{-3}$) plasma within the channel was sufficient to arrest catastrophic self-focusing when the laser beam was centimeter-scale. However, it was not clear what was the physical ionization mechanism and the density of electrons inside the channel.\par
As mentioned earlier, in the case of ionization by a LWIR laser ($\lambda$ = 10.6 $\mu$m, I = 10$^{12}$ W/cm$^2$, $\lambda_k \sim$ 0.7), the photon energy is so small that multi-photon ionization is near impossible. In addition to that, the estimated peak intensity in the channel of 10$^{12}$ W/cm$^2$ was well below the tunneling threshold for both N and O atoms. Therefore, cascade or avalanche ionization needs to be re-examined as the candidate mechanism for air breakdown even when ps pulses are used. At long wavelengths the generation of electrons needed to generate a collisional avalanche can happen if the kinetic energy in the oscillating laser field exceeds the ionization energy of O$_2$/N$_2$ molecules. For instance, at CO$_2$ laser wavelengths, the maximum oscillating energy of a free electron in the laser field with a peak intensity of 10$^{12}$ W/cm$^2$ is $\sim$50 eV---ideal for further collisional ionization of bound electrons \cite{Bro59}. As a result of such impact ionization two electrons appear in the gas and the process of cascading ensues that rapidly increases the free electron density. It is a well understood process from microwave and DC breakdown studies \cite{Mac66} and the number of electrons increases exponentially. As might be expected, the free electron density depends strongly on the laser pulse duration, explaining why the breakdown is expected to have a pulse length dependent threshold. After all, during the time of the laser pulse interaction with gas, the electron density must reach a certain critical value to completely screen the laser pulse and stop its propagation. In interaction of air-plasmas with a nanosecond CO$_2$ laser, observation of complete screening of pulse confirms the plasma density $\sim$10$^{19}$ cm$^{-3}$ was reached \cite{Kwo88,Gor99}. The origin of the first (seed) free electrons is still a puzzle. The best current hypothesis to this puzzle is either an ever-present low density of electrons due to a constant flux of cosmic rays \cite{Rai77} or enhanced ionization due to field-enhancement at the surface of aerosol (dust) particles \cite{Len73,Smi75}. The latter is a more likely explanation as the breakdown threshold is seen to dramatically increase when extremely pure gases are used to produce a breakdown with ns long pulses in free induction decay experiments \cite{Yab74b}.\par
The literature on the topic of laser spark evolution describes several mechanisms that can explain the backward propagation of breakdown, each occurring in various conditions. As mentioned earlier, from low to high laser intensities, these mechanisms are the laser-supported combustion wave (LSCW) \cite{Rai77}, laser-supported detonation wave (LSDW) \cite{Ram64,Rai77}, laser-supported radiation wave (LSRW) \cite{Rai77}, fast ionization wave (FIW) \cite{Shi19,Fis80}, and the breakdown wave \cite{Rai77}. However, several of these mechanisms follow a very similar pattern. At the beginning, the air is transparent to the laser radiation and therefore the initial breakdown point is usually at or in the proximity of the vacuum focus of the laser beam. As the plasma density increases via avalanche ionization, other mechanisms such as collisional absorption heat up the plasma and further increase the electron density as long as the pump pulse is present. As the plasma density nears the critical density ($\omega_p$ = $\omega_o$), plasma density gradients make it strongly refracting and eventually opaque. Here, $\omega_o$ is the laser frequency and $\omega_p=n_ee^2/m\varepsilon_o$ is the plasma frequency, where $n_e$ is the electron density, $e$ and $m$ are the electron charge and mass, respectively, and $\varepsilon_o$ is the permittivity of free space. This plasma continues to absorb more laser energy and provides both electrons and UV photons to assist in additional breakdown in the immediate upstream region \cite{Mey77}. This effect then cascades as energy is transferred from the ionized region to successive surrounding layers of air, heating them and causing additional ionization. Several mechanisms have been put forward to explain the observed velocity of the backward propagation of the plasma towards the laser. The main difference between different mechanisms is how the energy is transferred to the next upstream layer, whether by thermal conduction (LSCW), shockwaves (LSDW), or plasma radiation (LSRW). The FIW is similar to the LSRW in that the downstream plasma provides upstream gas layer seed electrons via photoionization by UV radiation (<100 nm), which then generate more electrons via the avalanche ionization process in the incoming laser field. Reported ionization front velocities in the literature \cite{Rai77} vary a lot, depending strongly on pulse lengths. Free running, millisecond Nd:Glass and ruby lasers produced $\sim$10$^4$ cm/s velocities by the LSCW mechanism after the initial spark is produced either on a pair of sharp point electrodes or with a 2$^{nd}$ initiating laser. Nanosecond, Q-switched ruby lasers reached the next regime of LSDW with $\sim$10$^7$ cm/s velocities. The fastest observed motion of $\sim$10$^8$ cm/s was achieved with a picosecond, Q-switched Nd:Glass laser in the FIW regime \cite{Alc68b}.\par
One exception to this group is the breakdown wave mechanism, which does not require any transfer of energy, but is instead a consequence of rising pulse power in a focusing geometry. In theory of the breakdown wave described in Ref.~\onlinecite{Rai77}, the electron density $n_e$ in a laser field increases exponentially from some initial value $n_0$ with an avalanche time constant $\theta$, such that the growth rate is proportional to laser intensity $I$,
\begin{subequations}
\label{eq1}
\begin{equation}
    n_e=n_0exp\left(\int_{0}^{t}\frac{dt'}{\theta}\right),\label{eq1a}
\end{equation}
\begin{equation}
    \theta^{-1}=aI(z,t),\label{eq1b}
\end{equation}
\end{subequations}
where $a$ is an unknown proportionality constant that will factor out later. We assume that the breakdown becomes visible when electron density reaches some value $n_1$. This will occur for different slices of the laser propagation channel along $z$ at different times $t$, governed by
\begin{equation}
    \int_{0}^{t}\frac{dt'}{\theta}=a\int_{0}^{t}I(z,t')dt'=\ln{\left(\frac{n_1}{n_0}\right)}=b,\label{eq2}
\end{equation}
where $b$ is another unknown constant---the number of collisional generations needed to reach n$_1$ from n$_0$. Now we write the laser intensity as a function of space and time,
\begin{equation}
    I(z,t)=\frac{P_m}{\pi[r(z)]^2}\phi(t),\label{eq3}
\end{equation}
where $P_m$ is the max laser pulse power, $r(z)$ is the radius of the optical channel at slice $z$, and $\phi(t)$ is a dimensionless pulse shape factor as a function of time. Substituting Eq.~\ref{eq3} into Eq.~\ref{eq2} gives
\begin{subequations}
\label{eq4}
\begin{equation}
    a\frac{P_m}{\pi r^2}\Phi(\tau)=b,\label{eq4a}
\end{equation}
\begin{equation}
    \Phi(\tau)=\int_{0}^{\tau}\phi(t')dt',\label{eq4b}
\end{equation}
\end{subequations}
where $\Phi(\tau)$ is the time-integrated pulse shape, up to time $\tau$. Now we take Eq.~\ref{eq4a} at the optical focus with radius $r_0$ and time-to-breakdown $\tau_0$, and divide it by the same equation taken at another point upstream (with larger $r(z)$ and unknown time $\tau$),
\begin{equation}
    \frac{\Phi(\tau_0)}{\Phi(\tau)}=\frac{r_0^2}{[r(z)]^2}=\left(\frac{r_0}{r_0+z\tan(\alpha)}\right)^2,
    \label{eq5}
\end{equation}
where $\alpha$ is the laser half convergence angle. Then the final equation for breakdown position $z$ as a function of time $\tau$ for the breakdown wave theory is 
\begin{equation}
    z(\tau)=\frac{r_0}{\tan(\alpha)}\left(\sqrt{\frac{\Phi(\tau)}{\Phi(\tau_0)}}-1\right).
    \label{eq6}
\end{equation}
\begin{figure}
\includegraphics[width=\columnwidth]{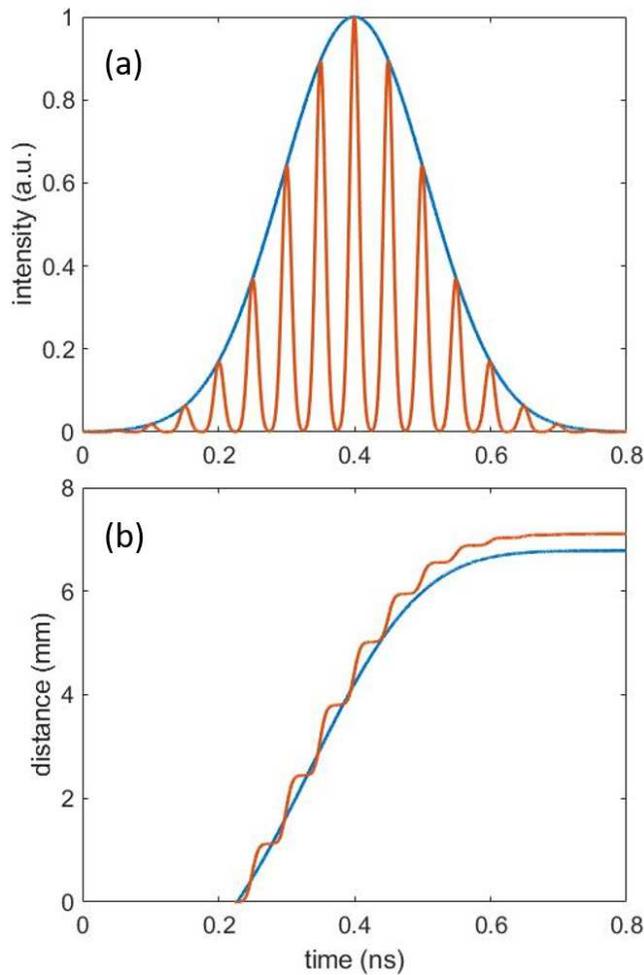}
\caption{\label{fig1}(a) Example pulse shapes: smooth envelope (blue) and pulse train (red). (b) Corresponding breakdown wave theory Eq.~\ref{eq6} for the two pulse shapes in a), using experimental parameters from Sec.~\ref{subsec4.2} and a fixed value for $\tau_0$.}
\end{figure}
In our case, we assume that there is a pulse train made up of 3 picosecond FWHM pulses each separated by 18.5 ps but we can nevertheless use this equation to calculate the breakdown position as a function of time and hence the propagation velocity of the breakdown wave with the experimental data discussed later in the paper. Example pulse shapes are given in Fig.~\ref{fig1}a to demonstrate why we can use the pulse envelope with Eq.~\ref{eq6} instead of having to use the fully modulated pulse train. Figure~\ref{fig1}b shows the qualitative result of applying Eq.~\ref{eq6} to both temporal profiles with a fixed value for $\tau_0$. The two cases show essentially the same curve (with some modulation for the pulse train case, red) because Eq.~\ref{eq6} is normalized by the denominator $\Phi(\tau_0)$ inside the radical. This normalization by the pulse integral allows us to approximate the true temporal shape with a much simpler smooth envelope. We will use $\tau_0$ as a fitting parameter later with the experimental results, because the time-to-breakdown in the focus fluctuates shot to shot. It should be noted that this envelope approximation for a pulse train only works because the temporal separation between pulses is shorter than the characteristic diffusion and recombination times of the newly born electrons. This way, the laser action of each pulse builds upon all of the pulses before it. The laser spark created with a train of mode-locked, picosecond, near-infrared pulses by Wang and Davis \cite{Wan71} exhibited different behavior. They saw a series of individual breakdown sites where the periodicity suggested that each breakdown was seeded by shockwaves from the one preceding it. This is not surprising since the pulse separation in a mode-locked train was 12 ns instead of only 18 ps in this experiment.
\section{\label{sec4}Experiments}
\subsection{\label{subsec4.1}CO$_2$ laser system}
\begin{figure*}[t]
\includegraphics[width=1.7\columnwidth]{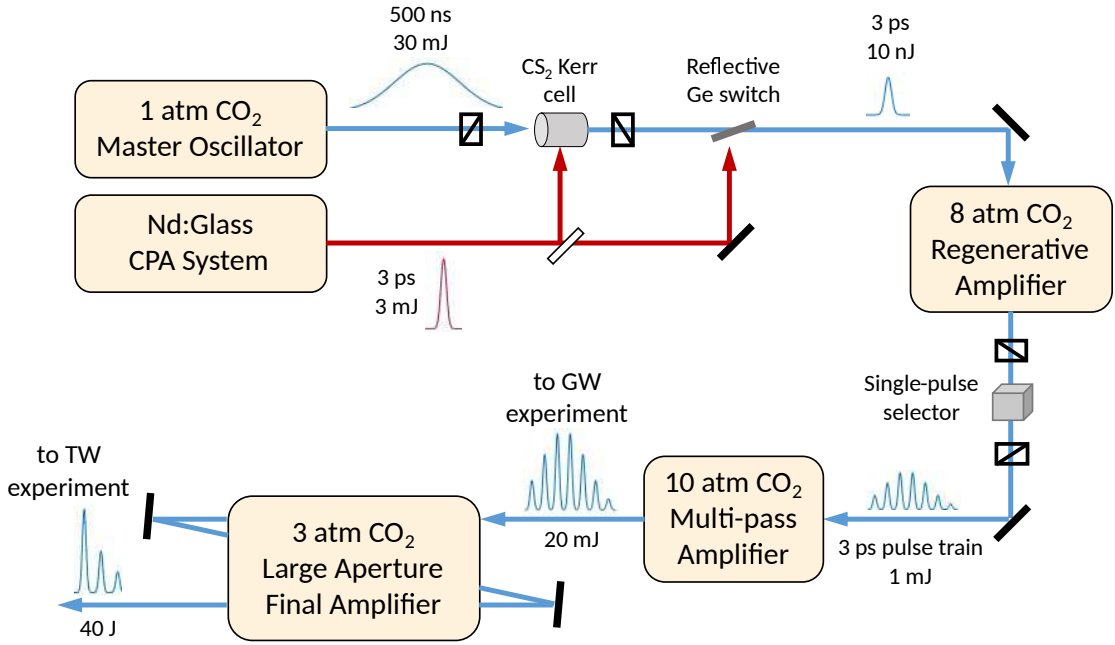}
\caption{\label{fig2}CO$_2$ master oscillator power amplifier system used for both the tightly focused laser experiment and the large diameter plasma channel experiment, without and with the single-shot final amplifier, respectively.}
\end{figure*}
Generation of high-power picosecond CO$_2$ laser pulses is realized in master oscillator-power amplifier (MOPA) systems in which a weak picosecond seed pulse is further amplified in a chain of multi-atmosphere CO$_2$ amplifiers \cite{Pig14}. In Fig.~\ref{fig2} we show schematically the UCLA CO$_2$ MOPA system used in the experiments. For both the tightly focused laser experiment (Sec.~\ref{subsec4.2}) and the large diameter plasma channel experiment (Sec.~\ref{subsec4.3}), the front end and first two stages of amplification were the same—-the only difference is that a 3-pass large-aperture CO$_2$ final amplifier was additionally used to reach TW power levels for the large diameter plasma channel experiment. First, to generate a broad bandwidth 10 $\mu$m seed pulse, we use a 3 ps Nd:Glass laser pulse for polarization gating of a much longer 10 $\mu$m pulse (from a typical transversely-excited atmospheric CO$_2$ laser \cite{Fil02}) using a liquid CS$_2$ cell. For additional laser contrast, we then reflect the 10 $\mu$m pulse off a cold plasma produced on the surface of a Ge polarizer at Brewster’s angle \cite{Cor85}. This produces a $\sim$3 ps but low energy ($\sim$few nJ) seed 10 $\mu$m pulse that is then further amplified by 10$^6$ in an 8 atm CO$_2$ regenerative amplifier. This amplification process inevitably results in the formation of a pulse train, called pulse splitting \cite{Toc12}, due to the incomplete spectral overlap of rovibrational lines in the CO$_2$ active medium. It is worth noting here that this is a picosecond pulse train, in addition to the nanosecond pulse train associated with the round-trip time of a regenerative amplifier cavity. We shall refer to the individual 3 ps pulses as micropulses within a single $\sim$150 ps macropulse, where the separation of micropulses is 18.5 ps, based on the inverse spectral spacing of rovibrational lines. We then select a single macropulse out of the regenerative amplifier with a half-wave CdTe Pockels cell placed between a polarizer and an analyzer, which also improves the nanosecond laser contrast. Energy is further increased via 2-pass amplification in a 10 atm CO$_2$ discharge module to GW peak power levels. As mentioned above, the TW experiment also requires 3-pass amplification in a 3 atm, CO$_2$ final amplifier operating in the power-broadened regime \cite{Toc12}. Note that GW power 10 $\mu$m pulses are generated at a repetition frequency of $\sim$1 Hz and TW power pulses are limited to one shot every 5-10 mins.\par
All experiments were done with a peak focused laser intensity of 10$^{12}$ W/cm$^2$ but with either a small focal spot obtained using a small f\# focusing optic or a large (cm diameter) beam that mimics the self-guided channels reported in Ref.~\onlinecite{Toc19}. The goal of the first experiment was to determine if the breakdown produced by a train of ps 10 $\mu$m laser pulses separated by 18 ps and containing only 20 mJ energy screens out the forward propagation of the pulse and induces a backwards propagating breakdown towards the focusing optic and if it does, what the velocity of propagation of this backward propagation is and determine the physical mechanism of the breakdown. The large diameter plasma channel experiment had a similar temporal structure as the tightly focused laser experiment but contained tens of Joules of energy. Such a large area beam can intercept many small dust particles floating in the air because of Brownian motion. The source of seed electrons in this case is breakdown that occurs on the surface of these individual particles. The question here is on what timescale do microscopic plasmas produced at these individual breakdown sites merge into one another to provide a macroscopic plasma that eventually recombines to produce a hot neutral channel that may be useful for other applications.\par
\subsection{\label{subsec4.2}Breakdown experiments using tightly focused laser beam}
\begin{figure*}[t]
\includegraphics[width=1.6\columnwidth]{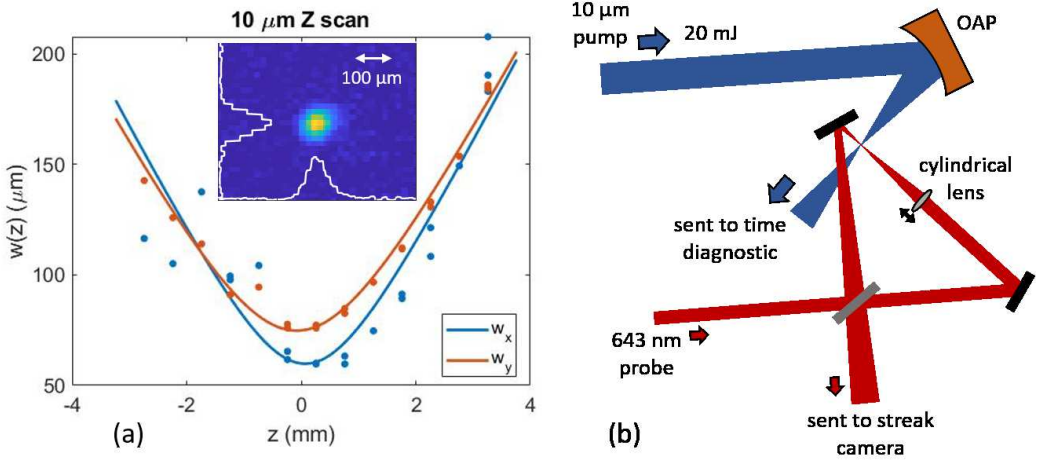}
\caption{\label{fig3}Tightly focused laser experimental setup with a Z scan of the pump beam focal geometry (a), a pyroelectric image of the 10 $\mu$m focal spot (inset), and Sagnac interferometer geometry (b).}
\end{figure*}
\begin{figure}[b]
\includegraphics[width=.9\columnwidth]{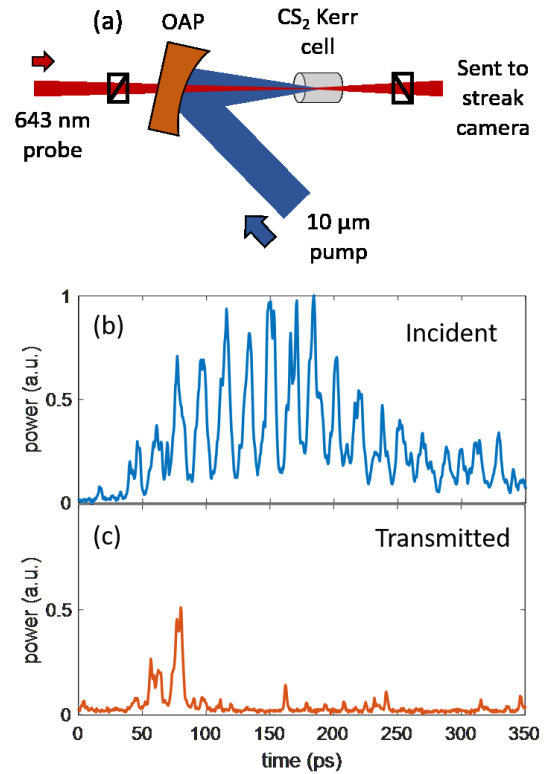}
\caption{\label{fig4}(a) Time diagnostic for upconverting 10 $\mu$m light to the visible for streak camera measurements using an off-axis parabolic (OAP) mirror with a hole in the center. Characteristic temporal measurements of the incident 10 $\mu$m pump pulse train (b) and transmitted light (c) when plasma breakdown is observed.}
\end{figure}

In this experiment, the $\sim$20 mJ, 10 $\mu$m macropulse was focused in laboratory air with an f=152 mm off-axis parabolic (OAP) mirror to generate a plasma spark. A scan of the beam profile was performed along the propagation path with a 5x beam-expanding telescope and a pyroelectric camera. The measured vertical and horizontal spot sizes of $\sim$70 $\mu$m are shown in Fig.~\ref{fig3}a with an image of the focal plane as an inset. We first studied evolution of the laser spark in space on relatively long timescales using a Michelson interferometer. This 2D transverse probing was done with a spatially-cleaned, 532 nm, 10 ns pulse from a frequency doubled Nd:YAG laser. Time-integrated interference fringes were captured with a charge-coupled device (CCD) camera providing a spatial resolution of $\sim$20 $\mu$m. To better understand the transverse expansion of the optical breakdown plasma, we measured how fast the CO$_2$ laser beam is screened by a near-critical and critical density plasma using a picosecond streak camera. For this purpose, the 10 $\mu$m light that was either transmitted beyond the laser-produced spark or refracted around it was recollimated and sent to a time-resolving diagnostic to measure the self-effect of the generated plasma on the pump pulse. This time-resolving diagnostic, shown in Fig.~\ref{fig4}a, consisted of upconverting the IR light to the visible spectrum via frequency mixing using another nonlinear Kerr cell filled with CS$_2$ and measuring the polarization-gated red light on a streak camera. Representative measurements for the incident and transmitted 10 $\mu$m radiation are shown in Fig.~\ref{fig4}b and Fig.~\ref{fig4}c, respectively. Note that the individual 3 ps pulses are not fully resolved in the figure.\par
To understand the fast ionization dynamics, the plasma spark was additionally probed transversely with a 643 nm diode laser inside the ring cavity of a Sagnac interferometer. This interferometer was chosen for its intrinsic phase stability as a common path interferometer \cite{Sch17}, where the reference and probe arms propagate in different directions around the ring cavity. For clarity, only the counter-clockwise propagating probe arm is shown in Fig.~\ref{fig3}b. A single cylindrical lens was placed exactly in the middle of the cavity to produce a horizontal line focus overlapping with the entire plasma spark while maintaining the same beam curvature for both interferometer arms. The cavity mirrors were tuned such that the counter propagating beams were combined with a zero-angle difference, producing a single bright fringe. Thus, any measured phase shift appears as a change in the fringe visibility or an amplitude change (from bright to less bright) rather than the more typical bending of fringes. The line focus was then image-relayed to a streak camera for a spatially- and temporally-resolved phase measurement of the 10 $\mu$m produced laser spark. In addition to probing, a Panasonic CCD camera was used to capture a transverse image of the plasma fluorescence (not shown in the figure).\par
\begin{figure*}[ht]
\includegraphics[width=1.8\columnwidth]{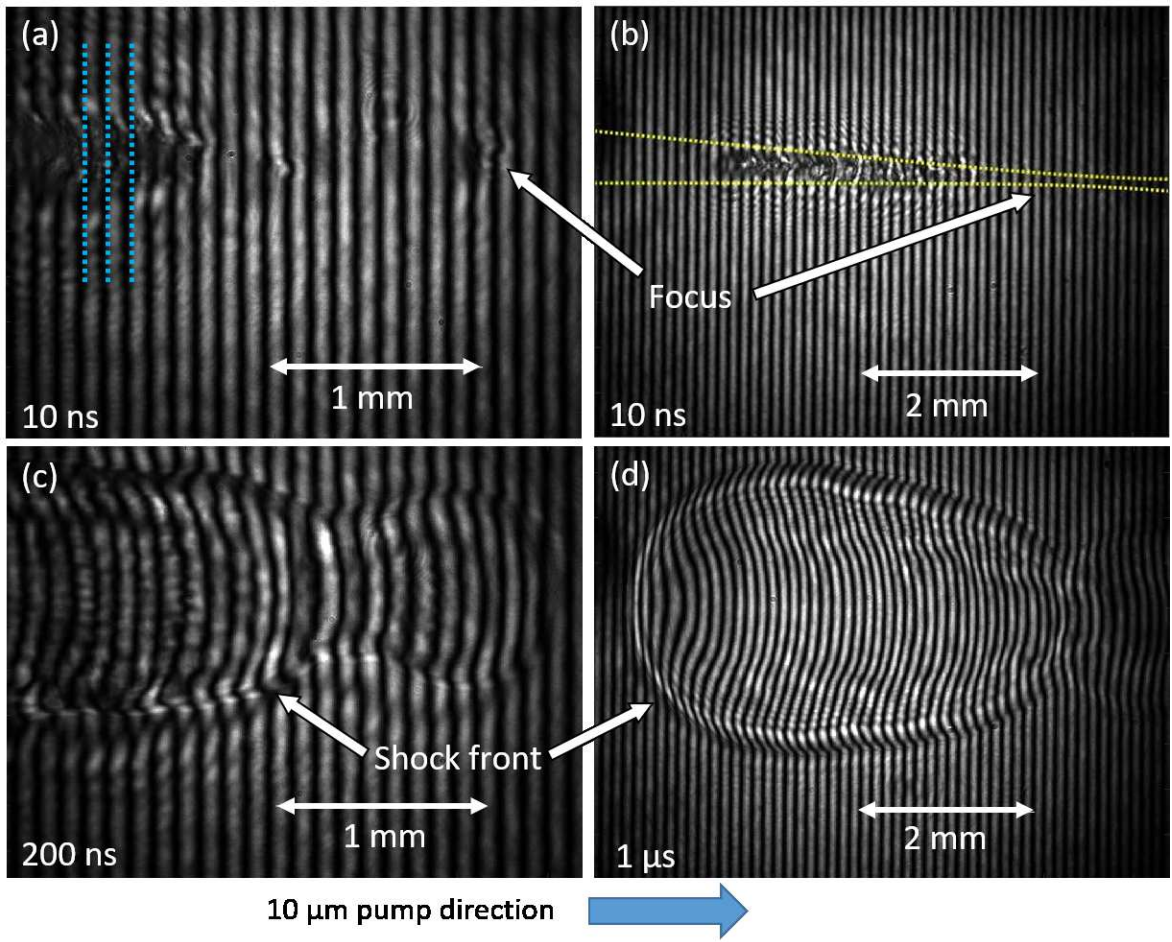}
\caption{\label{fig5}2D Michelson interference snapshots of the plasma spark and resultant shockwave evolution for characteristic laser shots. The vertical scale is the same as horizontal scale within each frame. Electron densities in the earlier frames (a and b) are $\geq$ 5x10$^{18}$ cm$^{-3}$. The blue dotted lines in a) help show the fringe bending and yellow dotted lines in b) are the pump laser diameter. Cylindrical shockwaves are visible in the later frames (c and d).}
\end{figure*}

2D interferograms obtained with the Michelson interferometer in Fig.~\ref{fig5} map out the plasma evolution from 10 to 1000 ns. In the earliest frames (Figs.~\ref{fig5}a and b both taken at @ 10 ns but with a factor of 2 different magnification) the plasma appears to be conical bounded by the cone angle of the laser intensity envelope. On the longer time scale (Figs.~\ref{fig5}c and d) we observe a more cylindrical expansion bounded by shockwaves. The early frames (Figs.~\ref{fig5}a and b) show the resultant plasma after the pump beam has passed through and the probe beam measures $\sim\pi/2$ rad phase shifts, corresponding to 5x10$^{18}$ cm$^{-3}$ electron densities. The later frames (Fig.~\ref{fig5}c taken at 200 ns and Fig.~\ref{fig5}d taken at 1 $\mu$s) are much more dominated by changes in neutral particle density associated with outward propagating shockwaves. This is evidenced by the change in the sign of the curvature of the fringes that occurs within the shockwave. We expect the shockwave itself to have a higher neutral density than the surrounding atmospheric gas, while the region enclosed by the shockwave would have a reduced density of hot neutral gas. Thus, the fringes bend one way for the shockwave and then bend back for the volume of reduced density. This is seen most clearly in Fig.~\ref{fig5}d with phase shifts up to $3\pi/2$. From the transverse expansion velocity of 2x10$^5$ cm/s, we get a temperature of $\sim10^4$ K assuming an average mass of 30 amu.\par
\begin{figure*}[!t]
\includegraphics[width=1.76\columnwidth]{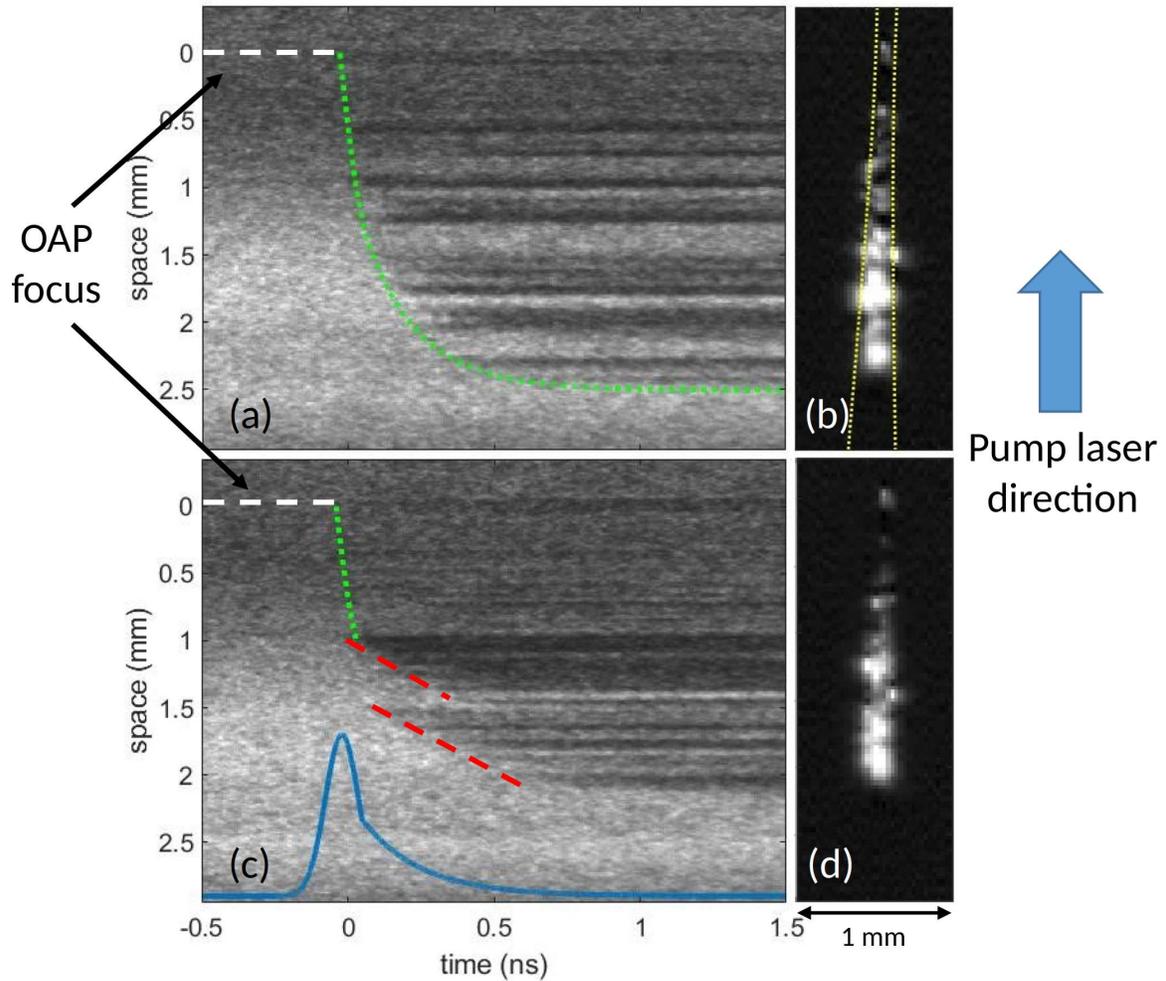}
\caption{\label{fig6}Spatiotemporal Sagnac interference measurements for two laser shots (a/b and c/d) with images of the plasma fluorescence (b,d) measured with a CCD camera. A majority of laser shots look like a) with a few exceptions that occurred in approximately 10 percent of the shots such as one shown in c) where simultaneous ionization of two aerosol (dust) particles took place. The white dashed lines (a,c) mark position of the best focus, green dotted lines (a,c) are breakdown wave theory curves using the solid blue line (c) smooth pulse envelope (a.u.) matched to the measured pulse train (Fig.~\ref{fig4}b), and red dashed lines (c) are examples of fast ionization wave. The yellow dotted lines in b) show the pump laser diameter. Time t=0 is set to where interference first appears.}
\end{figure*}

In order to better understand spatio-temporal dynamics of the optical breakdown plasma, we measured plasma evolution in the 0.01-2 ns window using a different diagnostic based on a fast streak camera as a readout device for a Sagnac interferometer in Fig.~\ref{fig3}b. Data from two characteristic laser shots is shown in Fig.~\ref{fig6}, with the streak interferograms (Figs.~\ref{fig6}a and c) shown alongside the plasma fluorescence images (Figs.~\ref{fig6}b and d) such that the vertical spatial axes line up within $\pm$ 0.1 mm. The 10 $\mu$m focal plane is at z=0 mm with positive numbers representing upstream of the focus. While the position of the vacuum best focus is well known, the t=0 or the arrival time of the structured laser pulse is not known and is arbitrary. What is clear is that once a sufficiently dense plasma is formed at the best focus position, the subsequent layers of the plasma are formed progressively later moving back towards the laser. The interferometric streaks show that the plasma generation is near-simultaneous for several localized points in space along the 10 $\mu$m propagation path as well as a more general trend of backward motion toward the laser source over subnanosecond timescales. While position of the first plasma breakdown site fluctuated a little, it always appeared within 50 $\mu$m of the vacuum focus of the pump laser. The dotted green lines show the best fit of the “breakdown wave” theory \cite{Rai77}, to be discussed later. We also occasionally saw several plasma beads appear in a very linear pattern travelling upstream toward the laser, as shown in Fig.~\ref{fig6}c with dashed red lines. The measured phase shifts > 1 rad correspond to electron densities > 6x10$^{18}$ cm$^{-3}$ in a good agreement with early time Michelson interferometry data. However, it should be noted that this is a lower-bound value since a faster change in the electron plasma density will simply not be distinguished by a streak camera.\par
We observed that for all laser shots with a peak intensity above 200 GW/cm$^2$, air was fully ionized and the plasma was formed in the area upstream from the focus. It should be noted that this value is significantly higher than the breakdown threshold of $\sim$1 GW/cm$^2$ reported for 100 ns long CO$_2$ laser pulses \cite{Smi75}. The small plasma in the focus, as seen in fluorescence images in Figs.~\ref{fig6}b and d, is followed by a dark zone and extends further upstream as a family of bright plasma beads expanding over $\sim$3 Rayleigh lengths. Such a beaded plasma is characteristic of optical breakdown due to avalanche ionization \cite{Ram64,Alc68a}. The fast transverse and longitudinal plasma expansion quickly cuts off transmission of the pump pulse, as shown in Fig.~\ref{fig4}c—--only the first few micropulses are transmitted. We measured an energy throughput ranging from 10 to 20\% on each laser shot. Thus, densities $\geq$ 10$^{19}$ cm$^{-3}$ are reached in the focus very early during the macropulse preventing downstream propagation and ionization of air. Also in Figs.~\ref{fig6}a and c, there is significant shot-to-shot fluctuation in specific position of plasma beads and corresponding dark lines from probe light experiencing at least a $\pi/2$ phase shift. The estimated electron density assuming $\sim$100 $\mu$m diameter plasma is $\sim$6x10$^{18}$ cm$^{-3}$.\par
\begin{figure}[ht]
\includegraphics[width=.95\columnwidth]{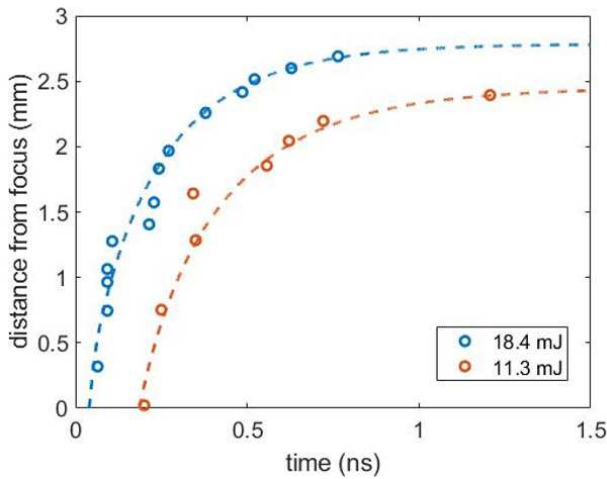}
\caption{\label{fig7}Location and timing of avalanche breakdown sites when they are first detectable with the Sagnac interferometer for maximum and minimum energy laser shots: high power (blue circles) and low power (red circles). Dashed curves are Eq.~\ref{eq6} with reasonable pulse train envelopes, as described in the text. Time t=0 is somewhat arbitrary for each dataset to within $\pm$ one micropulse (18 ps), and separated here for clarity. Note that these are different laser shots from Fig.~\ref{fig6}}
\end{figure}
It is very likely that all of the breakdown plasma propagation mechanisms contribute to the observed plasma dynamics. However, we find that in air breakdown initiated by a train of ps 10.6 $\mu$m laser pulses the breakdown wave mechanism is the most dominant because at 10$^{12}$ W/cm$^2$ intensities, air layers located at several Rayleigh lengths from the focus would be exposed to fields much above the optical breakdown threshold and simply ionized by the incoming laser micropulses. Figure~\ref{fig7} shows where and when high density plasma first becomes detectable by the Sagnac interferometer for the highest and lowest energy laser shots. The overall curve for each shot matches the breakdown wave theory Eq.~\ref{eq6} reasonably well when we use a pulse envelope similar to our measured pulse train (Fig.~\ref{fig4}b) for $\phi(t)$. Although there are several envelope shapes that can be used to match the plateau seen in the data of Fig.~\ref{fig7}, the most important part is the rising edge of the pulse train that defines the initial backward velocity. We also use $\tau_0$ as a fitting parameter (approximately 150 ps in the experiment), which fixes how much time-integrated energy flux was needed to reach breakdown densities. Then Eq.~\ref{eq6} maps out where and when this value is reached for the entire pulse train envelope. Figure~\ref{fig6} has green dotted lines for Eq.~\ref{eq6} on top of the raw interferograms. A simplification that Raizer \cite{Rai77} made was the assumption of right-angle triangular pulses, giving a breakdown velocity of
\begin{equation}
    v_{br}=\frac{r_0}{\tau_0\tan(\alpha)}.\label{eq7}
\end{equation}
For our experimental values, Eq.~\ref{eq7} gives velocities > 10$^9$ cm/s, which agrees with experimental data during the rise time of the macropulse. We also see a role played by a second mechanism that causes backward propagation with velocities > 10$^8$ cm/s only appearing after the peak of the laser pulse (red dashed lines in Fig.~\ref{fig6}c), possibly seeded by dust particles as their appearance is very random in nature. In the experiment, such accidental simultaneous ionization due to aerosol (dust) particles occurred approximately 10\% of laser shots. Our main finding is that when ps rather than ns 10 $\mu$m laser pulses are used, the backward propagation velocities are 10 to 20 times faster than can be explained by the I$^{1/3}$ scaling associated with the LSDW \cite{Ram64,Rai77}. In addition, the FIW should dominate LSRW at pressures over 0.1 atm \cite{Shi19}, as in our atmospheric experiment, because avalanche will take over at higher pressures after it is seeded by UV photons. The FIW mechanism seems to dominate plasma expansion in the transverse direction. Here, as measured by the second streak camera diagnostic (see Fig.~\ref{fig4}c), the self-screening of CO$_2$ laser pulses occurs over $\geq$ 20 ps defined by the separation between micropulses. Note, that both streak camera diagnostics acquired the data on the same laser shot to minimize effect of shot-to-shot variations. The deduced velocity around 5x10$^8$ cm/s is too fast for the LSDW mechanism \cite{Ram64} and can only be explained by photoionization of the focal volume by UV light (consistent with a FIW) of the primary plasma generated in the very focus.
\subsection{\label{subsec4.3}Large diameter plasma channel experiment}
\begin{figure*}[p]
\includegraphics[width=1.65\columnwidth]{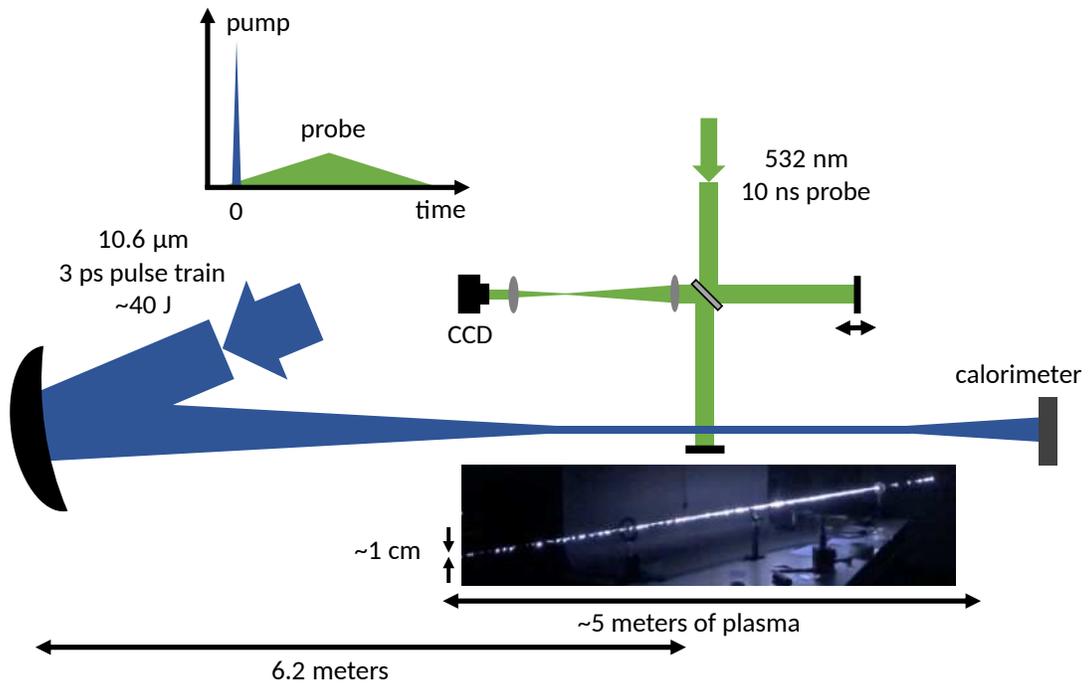}
\caption{\label{fig8}Large diameter plasma channel experimental setup and plasma fluorescence picture.}
\end{figure*}
\begin{figure*}[p]
\includegraphics[width=1.7\columnwidth]{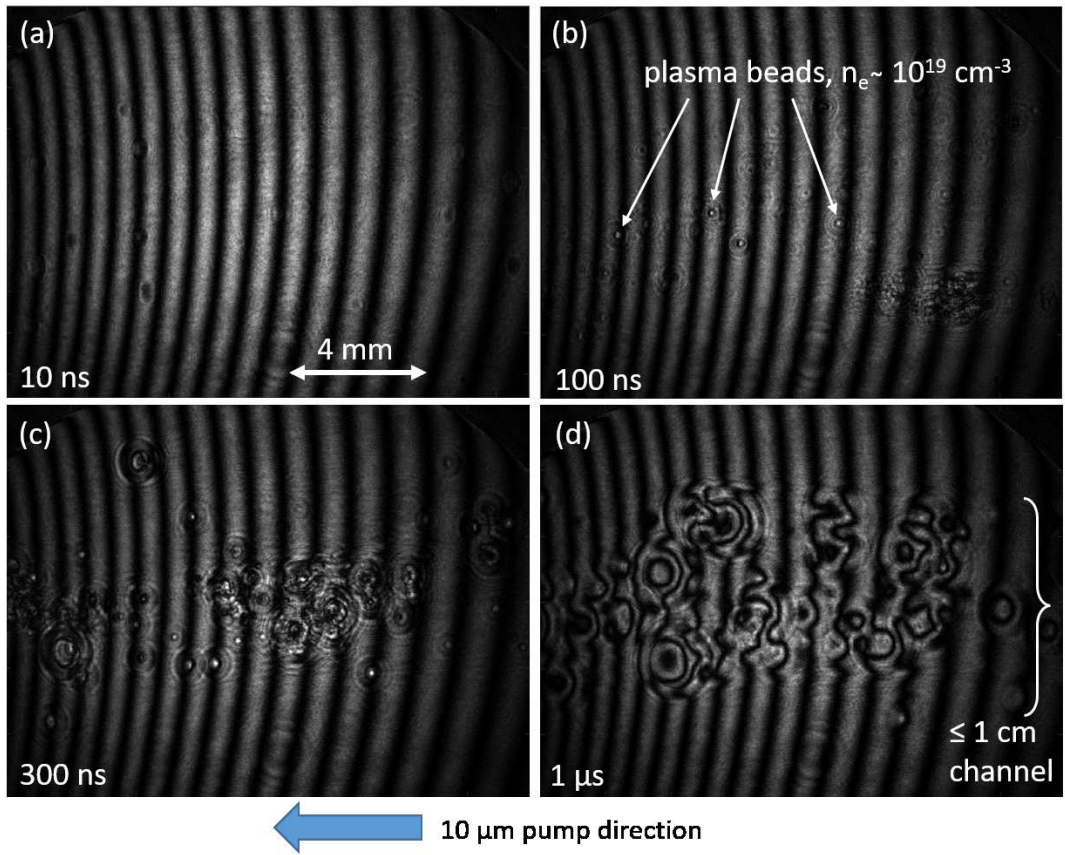}
\caption{\label{fig9}2D interference snapshots of the beaded plasma structure and resultant shockwave evolution for characteristic laser shots. The vertical and horizontal scales are the same in all frames. Phase shift is on the threshold of detection for the earliest time of 10 ns (a). Plasma beads and spherical shockwaves are seen together for 100 and 300 ns (b and c). Phase shift has fallen below detection threshold by 1 $\mu$s as shockwaves overlap (d).}
\end{figure*}
For the second experiment, we amplified the 20 mJ pulse train further in a 3-pass, multi-atmosphere, CO$_2$ final amplifier operating in a power-broadened regime. With a final energy of $\sim$40 J, the 76 mm diameter beam was focused with an R=14.5 m curved copper mirror in air down to a $\sim$1 cm (FWHM) spot size. Thereafter the laser propagated at least 5 meters producing a visible large diameter plasma column as shown in Fig.~\ref{fig8}. It should be noted that even though we observed a measurable shift of the nonlinear focus due to Kerr self-focusing, a high-density plasma channel effectively screened almost all laser energy and no channeling was observed behind the long plasma. We estimate that peak intensity of between 1-2 TW/cm$^2$ was reached, similar to the measurements in the previous section but in a much larger beam size. For each shot the energy throughput was measured with a large diameter calorimeter after the end of the visible plasma channel. The initial CO$_2$ laser beam evolved in several beamlets (filaments) containing $\sim$10\% of the initial total pulse energy. This beam structure was seen with the exposure of photo emulsion paper to be random and strongly affected by laser interaction with a near-critical density plasma of optical breakdown in air. Figure~\ref{fig8} shows the transverse probing setup of a Michelson interferometer using a 532 nm, 10 ns pulse. For such large diameter channels, one would expect that many dust particles in the laboratory air will be intercepted by the propagating CO$_2$ laser beam and ionized. The CCD captured 2D interferograms of the plasma near the nonlinear focus of the laser beam where the most plasma beads appeared.\par
\begin{figure}[t]
\includegraphics[width=.95\columnwidth]{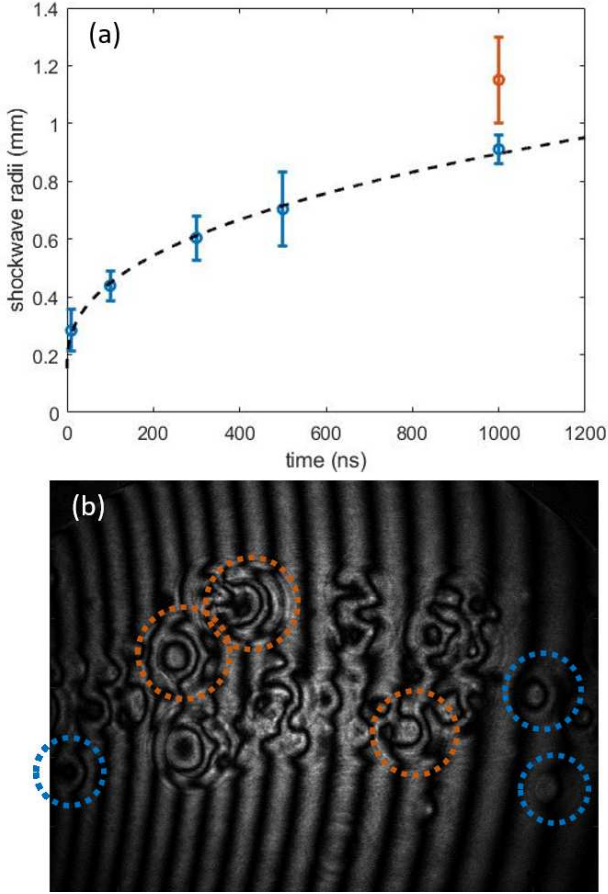}
\caption{\label{fig10}(a) Measured shockwave growth over time, fit to curve of Taylor's blast wave theory. Blue data are isolated shockwaves and red data point is overlapping shockwaves with some examples given in (b) for 1 $\mu$s. Error bars indicate the spread in data, as the measurement uncertainty was very small.}
\end{figure}

Some characteristic interferograms from different laser shots of the beaded plasma structure and how they evolve over a microsecond timescale are shown in Fig.~\ref{fig9}. Initially, the plasma formed on these dust particles has a very small size and therefore the phase shift associated with the density length product is below our detection threshold. By 100 ns, we can see bright centers inside of rings that continue to expand at later times. These bright spots have phase shifts > $\pi$ radians, corresponding to an initial electron density of > 10$^{19}$ cm$^{-3}$, and eventually fade away as the plasma recombines and falls below our detection threshold by the last frame at $\sim$1 $\mu$s. Later time frames (not shown) are dominated by “hot gas” dynamics due to recombination of the air-plasma \cite{Cam06}. These plasma beads appeared with an average density of 40 cm$^{-3}$ at the brightest part of the plasma channel with decreased density on either end, as seen in the picture in Fig.~\ref{fig8}. The rings in Fig.~\ref{fig9} are spherical shockwaves of disturbed neutral air particles that expand at velocities > 10$^5$ cm/s and eventually coalesce with one another on microsecond timescales, producing a large $\leq$ 1 cm channel of heated gas. The shockwave radii are plotted as a function of time in Fig.~\ref{fig10}a with a curve for Taylor’s blast wave theory \cite{Ram64,Tay50} fit to the data. The blue data points are for isolated shockwaves extracted for 10 local plasma beads, whereas the red data is from shockwaves that had much more overlap with each other at the longest time that they’re still measurable (examples are shown in Fig.~\ref{fig10}b). The estimated expansion velocity is initially five times the speed of sound (C$_S$) in air at STP, but quickly drops to 1.3C$_S$ by 500 ps. Detection of a very loud sound wave in the experiment supports the observation of supersonic shocks in air. Lastly, the laser energy throughput fluctuated around 10\% of the maximum energy with a vast majority of the energy being absorbed and scattered by the plasma.\par
The fundamental difference between the two experimental arrangements is the size of the laser beam, and hence the volume of air is exposed to the TW/cm$^2$ fields. The first experiment had some probabilistic chance of catching a dust particle that could seed avalanche breakdown \cite{Len73}, whereas the large spot size experiment was guaranteed to intercept several that in turn dominated the ionization process. Each initiation site undergoes cascaded (avalanche) ionization driven by the strong laser field until the air locally nears full single (outermost) electron ionization. The high-density plasma bead also becomes opaque to the laser because 10$^{19}$ cm$^{-3}$ is the critical density for 10 $\mu$m radiation. On the timescale of atomic motion, this laser energy is deposited nearly instantaneously and over a very small volume. The result is that each plasma bead launches its own spherical detonation wave. One can estimate the shockwave dynamics using the classical Taylor theory \cite{Tay50}. The shockwave displacement $x$ follows the equation
\begin{equation}
    x=\left(\frac{W}{\rho}\right)^{1/5}t^{2/5},\label{eq8}
\end{equation}
where $W$ is the deposited energy, $\rho$ is the mass density of the ambient gas, and $t$ is time. Eq.~\ref{eq8} comes from Taylor’s blast wave theory \cite{Tay50} and was originally meant for studying ultrafast atmospheric detonations where a large amount of energy is deposited in a small volume. However, it has been applied before with shockwaves created by a laser spark in air \cite{Ram64}. In our case blue data points correspond to 10-15 measured local shocks that are not overlapped with each other. There is a remarkable fitting of the experimental data with the Taylor theory for plasmas with the same temperature. However, for overlapping spacial areas of “hot gas islands” appearing at later times, there is a departure from the initial Taylor’s dependence (shown by red data point). The departure of this data can be explained by extreme overlap of the intersecting shockwaves that additionally heat gas around each other. The measured velocity of $\sim$2x10$^5$ cm/s is very similar to that seen in the tightly-focused geometry plasma expansion shown in Fig.~\ref{fig5}. Evolution of a plasma channel into a hot gas channel occurs through a merging of initial small-scale shocks driven by the plasma beads formed on aerosol particles. Although the shockwaves are not visible at longer times, we observed that the heated gas channel lasts through millisecond timescales due to the slow thermal diffusion process in air.
\section{\label{sec5}Conclusions}
We have presented interferometric data on plasma generation in atmospheric air by LWIR radiation at TW/cm$^2$ laser intensities. The ionization physics is dominated by the avalanche mechanism. For tight focusing geometries, backward plasma motion is observed and explained by a combination of the breakdown wave and fast ionization wave mechanisms. The measured velocities of the order of 10$^9$ cm/s are more than 10x that reported for the air breakdown with nanosecond CO$_2$ laser pulses. Another conclusion is that for a large diameter channel, relevant for future nonlinear self-guiding experiments, gas impurities such as aerosol (dust) particles play a very important role as they can seed the avalanche process. We observed breakdown with our $\sim$200 ps pulse trains on every shot with a peak laser intensity > 200 GW/cm$^2$, which is a much higher threshold than that observed with nanosecond and longer pulses. Plasma densities > 10$^{18}$ cm$^{-3}$ are reached within 10s of picoseconds, ultimately limiting the pulse length of any high power LWIR laser propagating through the atmosphere. At intensities above 200 GW/cm$^2$ the optical breakdown of air will eventually prevent long distance propagation of $\geq$ 100 ps CO$_2$ laser pulses in air. However, it is expected that this trend of higher breakdown thresholds for shorter pulses should continue. The nonlinear propagation of a few picosecond long, single LWIR pulse through atmospheric air demonstrated recently \cite{Toc19} indirectly indicates that the breakdown threshold can be further increased and its physics needs to be studied further.

\begin{acknowledgments}
This work was supported by the Air Force Office of Scientific Research grant FA9550-16-1-0139, Office of Naval Research Multidisciplinary University Research Initiative grant N00014-17-1-2705, and Department of Energy grant DE-SC0010064.
\end{acknowledgments}

\section*{Disclosures}
The authors declare no conflicts of interest.

\section*{Data Availability Statement}

The data that support the findings of this study are available from the corresponding author upon reasonable request.

\section*{References}
\nocite{*}
\bibliography{references}

\end{document}